\documentclass[sigchi]{acmart}
\renewcommand\footnotetextcopyrightpermission[1]{}
\usepackage{balance}

\begin{document}

%

%
\title{A Simple Deep Personalized Recommendation System}

%

\author{Pavlos Mitsoulis-Ntompos}
\authornote{Equal contribution to this research.}
\email{pntompos@expediagroup.com}
\author{Meisam Hejazinia}
\authornotemark[1]
\email{mnia@expediagroup.com}
\author{Serena Zhang}
\authornotemark[1]
\email{shuazhang@expediagroup.com}
\affiliation{%
  \institution{Vrbo, part of Expedia Group}
}

\author{Travis Brady}
\email{tbrady@expediagroup.com}
\affiliation{%
  \institution{Vrbo, part of Expedia Group}
}

%

\renewcommand{\shortauthors}{Mitsoulis-Ntompos, Hejazinia and Zhang, et al.}

\renewcommand\footnotetextcopyrightpermission[1]{} 

%

\begin{abstract}
Recommender systems are critical tools to match listings and travelers in two-sided vacation rental marketplaces. Such systems require high capacity to extract user preferences for items from implicit signals at scale. To learn those preferences, we propose a Simple Deep Personalized Recommendation System to compute travelers' conditional embeddings. Our method combines listing embeddings in a supervised structure to build short-term historical context to personalize recommendations for travelers. Deployed in the production environment, this approach is computationally efficient and scalable, and allows us to capture non-linear dependencies. Our offline evaluation indicates that traveler embeddings created using a Deep Average Network can improve the precision of a downstream conversion prediction model by seven percent, outperforming more complex benchmark methods for online shopping experience personalization.
\end{abstract}

%

%

\maketitle

\section{Introduction}
Personalizing recommender systems is the cornerstone for two-sided marketplace platforms in the vacation rental sector. Such a system needs to be scalable to serve millions of travelers and listings. On one side, travelers show complex non-linear behavior. For example, during a shopping cycle travelers might collect and weight different signals based on their heterogeneous preferences across various days, by searching either sequentially or simultaneously. Furthermore, the travelers might forget and revisit items in their consideration set \cite{de2012testing,chade2017sorting}. On the other side, marketplace platforms should match each of the travelers with the most personalized listing out of millions of heterogeneous listings. Many of these listings have never been viewed by any traveler or have only been recently onboarded, imposing data sparsity issue. In addition, the context of each trip might be different for travelers within and across different seasons and destinations (e.g. winter trip to mountains with friends, summer trip to the beach with family, etc.). Moreover, such a personalized recommender system should always be available and trained based on the most relevant data, allowing quick test-and-learn iterations, adapting to ever changing requirements of business. This personalized recommender system should suggest handful relevant listings to the millions of travelers visiting site pages (e.g. home page, landing page, or listing detail page), travelers receiving targeted marketing emails, or travelers faced cancelled bookings due to various reasons. 

To develop such a recommender system we need to extract travelers' preferences from implicit signals of their interactions using machine learning or statistical-economics models. Given the complexity and scale of this problem, we require high capacity models. While powerful, high-capacity models frequently require prohibitive amounts of computing power and memory, particularly for big data problems. Many approaches have been proposed to learn item embeddings for recommender systems \cite{wang2019survey,liang2016factorization,bogina2017incorporating,caselles2018word2vec}, yet learning travelers' preferences from those listing embeddings at scale is still an open problem. Indeed, such a solution needs to capture traveler heterogeneity while being generic and robust to cold start problems. We propose a modular solution that learns listings and traveler embeddings non-linearly using a combination of shallow and deep networks. We used down-funnel booking signals, in addition to implicit signals (such as listing-page-view), to validate our extracted traveler embeddings. We deployed this system in the production environment. We compared our model with three benchmark models, and found that adding these traveler features to the extant feature set in the already-existing Traveler Booking Intent model can add significant marginal values. Our finding suggests that this simple approach can outperform LSTM models, which have significantly higher time complexity. In the next sections we review related work, explain our model, review the results, and conclude. 

\section{Related Works}
Representation learning has been widely explored for large-scale session-based recommender systems (SBRS), \cite{grbovic2015commerce,lake2019large,wang2019survey}, among which collaborative filtering and content-based settings are most commonly used to generate user and item representations \cite{liang2016factorization,grbovic2015commerce,sedhain2015autorec}. Recent works have addressed the cold start and adaptability problems in factorization machine and latent factor based approaches \cite{mnih2008probabilistic,johnson2014logistic,wu2018session}. Other works have employed non-linear functions and neural models to learn the complex relationships and interactions over users and items on e-commerce platforms \cite{wu2018session,lake2019large}. In particular, word2vec techniques with shallow neural networks \cite{mikolov2013distributed} from the Natural Language Processing (NLP) community have inspired authors to generate non-linear entity embeddings \cite{grbovic2015commerce} using historical contextual information. State-of-the-art methods have used attention neural networks to aggregate representations in order to focus on relevant inputs and select the most important portion of the context \cite{chaudhari2019attentive}. Attention has been found effective in assigning weights to user-item interactions within the encoder-decoder and Long Short Term Memory (LSTM) architectures and collaborative filtering framework, capturing both long and short term preferences \cite{1904.05985,eide2018deep,lake2019large}. Similar to the spirit of our work, recent studies suggested simple neural networks, showing promising results in terms of performance, computational efficiency and scalability \cite{iyyer2015deep,arora2016simple,zhu2018learning}.\\

\section{Architecture and Model}

In this section, we will describe our model, which is based on the session based local embedding model. Our model has two modular stages. In the first stage, we train a skip-gram sequence model to capture a local embedding representation for each listing, we then extrapolate latent embeddings for listings subject to the cold start problem. In the second stage, we train a Deep Average Network (DAN) stacked with decoder and encoder layers predicting purchase events to capture a given traveler's embedding or latent preference for listings embedding. We also mention a couple of alternatives we evaluated for traveler embeddings. We denote each listing by $x_i$, so each traveler session $s_k(t_j)$ is defined as a sequence like $x_1, x_2,...$ for traveler $t_j$. We denote booking event conditional on listings recently viewed by the traveler with $b_k(t_j| x_{j1}, x_{j2},, .., x_{jt})$. Our contribution in this paper is mainly the second stage which we validate using a downstream shopping funnel signal. 
\subsection{Skip-gram Sequence Model}
The skip-gram model \cite{mikolov2013distributed} in our context attempts to predict listings $x_i$ surrounded by listings $x_{i-c}$ and $x_{i+c}$ viewed in a traveler session $s_k$, based on the premise that traveler's view of listings in the same session signals the similarity of those listings. We use a shallow neural network with one hidden layer with lower dimension for this purpose. The training objective is to find the listing local representation that specifies surrounding most similar manifold. More formally the objective function can be specified by the log probability maximization problem as follows:

$$\frac{1}{S}\sum_{s=1}^S\sum_{-c\leq j \leq c, j\neq 0} \log p(x_{i+j}|x_i)$$

where $c$ is the window size representing listing context. The basic skip-gram formulation defines  $p(x_{i+j}|x_{i})$ using softmax function as follows:

$$p(x_{i+j}|x_{i}) = \frac{ \text{exp}(\nu_{x_{i+j}} ^T \nu _{x_{i}} ) }{\sum_{x=1}^X \text{exp}(\nu_{x} ^T \nu_{x_{i}} ) }$$

where $ \nu_{x}$ and $\nu_{x_{i}}$ are input and output representation vector or neural network weights, and $X$ is the number of listings available on our platform. To simplify the task, we used the sigmoid formula, which makes the model a binary classifier, with negative samples, which we draw randomly from the list of all available listings on our platform. Formally, we use the following formula:
$p(x_{i+j}|x_{i}) = \frac{ exp(\nu_{x_{i+j}} ^T \nu _{x_{i}} ) }{1 + \text{exp}(\nu_{x_{i+j}} ^T \nu _{x_{i}} ) }$
for positive samples, and the following formula for negative ones:
$p(x_{i+j}|x_{i}) = \frac{1}{1 + \text{exp}(\nu_{x_{i+j}} ^T \nu _{x_{i}} ) }$.

We have two more issues to address, sparsity and heterogeneity in views per item. It is not uncommon to observe long tail distribution of views for the listings. For this purpose we leverage approaches mentioned by \cite{mikolov2013distributed} 
wherein especially frequent items are downsampled
using the inverse square root of the frequency. Additionally, we removed listings with very low frequency. 
To resolve the cold start issue, we leverage the contextual information that relates destinations (or search terms) to the listings based on the booking information. Formally, considering that the destinations $d_1, d_2, ..., d_D$ are driving $p_{id_1}, ..., p_{id_D}$, proportion of the demand for a given listing, we form the expectation of the latent representation for each location using $\nu_d=\frac{1}{N}\sum_{l=1}^Lp_{ld}\nu_{x_l}$, where $N$ is the normalizing factor and $L$ is the total number of destinations. Then, given latitude and longitude of the cold listing (for which we have no data), we form the belief about the proportion of demand driven from each of the search terms $p_{jd_1}, ..., p_{jd_D}$. Then, we use our destination embedding from the previous step to find the expected listing embedding for the cold listing as follows $\nu_{x_j}=\sum_{d=1}^Dp_{jd}\nu_{d}$. 

\subsection{Deep Average Network and Alternatives} 
In the second stage, given the listing's embedding from the previous stage we model traveler embeddings using a sandwiched encoder-decoder non-linear Relu function. In contrast to relatively weak implicit view signals, in this stage we leverage strong booking signals as a target variable based on historical traveler listing interaction. We have various choices for this purpose including Deep Average Network with Auto-Encoder-Decoder, Long Short Term Memory (LSTM), and Attention Networks. The simplest approach is to take the point-wise average of the embedding vector and use it directly in the model. The second approach could be to feed the average embedding into a dimensionality expansion and reduction non-linear encoder-decoder architecture, or Deep Average Network to extract the signals \cite{iyyer2015deep}. The third approach could incorporate LSTM network \cite{lang2017understanding,Sheil2018PredictingPI}, testing the hypothesis that the traveler signals information that they gathered by looking at different listings in the shopping funnel. The fourth approach could have an attention layer on the top of LSTM \cite{zhou2016attention}, hypothesizing that they allocate different weights on various latent features before their booking.

We take a probabilistic approach to model traveler booking events $\mathrm{P}(Y_j)$ based on the embedding vectors of historical units they have interacted with $\nu_{j1},, .., \nu_{jt}$. Formally, given the traveler embeddings (or last layer of the traveler booking prediction neural network $f(\nu_{j.})$), the probability of the booking is defined as:

\begin{equation}
\mathrm{P}(Y_j|\nu_{j1}, \nu_{j1},, .., \nu_{jt}) = \text{sigmoid}(f(\nu_{j.}))
\end{equation}

where, the Deep Average Network layers and $f$ are defined as:

\begin{align}
f(\nu_{j.}) &= \text{relu}(\omega_1 \cdot h_2(\nu_{j.}) + \beta_1)\\
h_1(\nu_{j.}) &= \text{relu}(\omega_2 \cdot h_1(\nu_{j.}) + \beta_2)\\
h_2(\nu_{j.}) &= \text{relu}(\omega_3 \cdot \frac{1}{k}\sum_{i=1}^{t}{\nu_{ji}}) + \beta_3)
\end{align}

Alternatively, we can use an LSTM network with forget, input, and output gates as follows:

\begin{multline}
f(\nu_{j}^{t}) = \text{sigmoid}(\omega_f [h_t, \nu_j^t]+\beta_f)
\cdot f(\nu_{j.}^{t-1})\\
+ \text{sigmoid}(\omega_i[h_t, \nu_j^t]+\beta_i)\cdot \text{tanh}(\omega_c[h_{t-1},\nu_j^t]+\beta_c)
\end{multline}

And finally, we can also use an attention network on the top of LSTM network as follows:
\begin{equation}
f(\nu_{j}) = \text{softmax}(\omega^T \cdot h_{T})\text{tanh}(h_{T})
\end{equation}
where $\omega_., \beta_.$ are weight and bias parameters to estimate and $h_t$ represents the hidden layer parameter or function to estimate.

\begin{figure}
  \includegraphics[width=\linewidth]{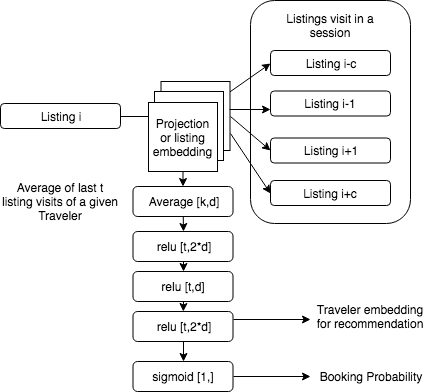}
  \caption{Deep Average Network (DAN) on the top of skip-gram network.}
  \label{fig:skipDAN}
\end{figure}

Among these models, DAN is more consistent with Occam's razor principle, so it is more parsimonious, and faster to train. However, LSTM and Attention Networks on the top of it are more theoretically appealing. As a result, from the pragmatic stand point, for millions of listings and travelers DAN seems to be more appealing for deployment as depicted in Figure \ref{fig:skipDAN}.

We use adaptive stochastic gradient descent method to train the binary cross entropy of these neural networks. The last question to answer is how are we planning to combine the traveler and listing embedding for personalized recommendations. This is a particularly challenging task as traveler embeddings is non-linear projection of listings embedding with a different dimension. As a result, they are not in the same space to compute cosine similarity. We have various choices for this solution, including approaches such as factorization machine and svm with kernel that allow modeling higher level interactions at scale. We defer the study of this approach to our next study. 

\section{Datasets}
For the experiments, anonymized clickstream data is collected for millions of users from two different seven-day periods. Specifically, the click stream data includes user views and clicks of listing detail page logs, search requests, responses, views and clicks logs, homepage views and landing page logs, conversion events logs, per visitor and session. The first click-stream dataset was used to generate embeddings using Deep Average Network and the LSTM with Attention. The second click-stream dataset was used to evaluate the learned embeddings on the Traveler Booking Intent Model. We split each of the data sets into train and test set by 70:30 proportion randomly, based on users. In other words, users that are in the train set are excluded from the test set, and vice versa. 

\section{Experiments and Results}

In this section we describe the experimental setup, and the results obtained when comparing the accuracy uplift of our Deep Average Network based approach to various baselines on a downstream conversion prediction model. The Traveler Booking Intent XGBoost model is such a downstream model. It is trained using LightGBM \cite{LightGBM} and uses a rich set of hand-crafted historical session-based product interaction features in order to predict the booking intent probability\footnote{We call it booking intent as our model predicts booking request from travelers, which needs a couple of steps to be confirmed as booking.}. In order to evaluate offline our proposed methodology, we concatenated the hand-crafted features with the traveler embeddings, generated by all different model settings.

The three baseline methods that we compare against our proposed Deep Average Network on the top of Skip-Gram include the following:
\begin{enumerate}
  \item \textbf{Random}: a heuristic rule that chooses a random listing embedding, among those listings a traveler has previously interacted with, in the current session.
  \item \textbf{Averaging Embeddings}: a simple point-wise averaging of listing embeddings a traveler has previously interacted with, in the current session.
  \item \textbf{LSTM with Attention}: A recurrent neural network, inspired by \cite{lang2017understanding,10.1007/978-3-030-10997-4_9,Sheil2018PredictingPI}, that uses LSTM units and an attention mechanism on top of it in order to combine embeddings of listings a user has previously interacted with, in the current session.
\end{enumerate}

\balance

\subsection{Results}
We ran our training pipeline on both CPU and GPU production systems using Tensorflow \cite{tensorflow2015-whitepaper}. We cleaned up the data using Apache Spark \cite{Zaharia2016ApacheSA}, and the input data to training pipeline had observations from millions of traveler sessions. The training process for LSTM models typically took 3 full days of time, while training DAN took less than 8 hours on CPU. Given that our recommender system needs to be iterated fast for improvement and infer in real-time with high coverage, DAN model scales better. Moreover,  we modified the cost function to give more weight to minority class (i.e. positive booking intent)  in order to combat the imbalanced classes in the data sets.

We evaluated the performance of the Traveler Booking Intent model on the different settings using the test data set based on AUC, Precision, Recall and F1 scores. The best results of each model are shown in Table \ref{tab:comparison}. It shows that our proposed Deep Average Network approach contributes more uplift to the downstream Traveler Booking Intent model.

\begin{table}[H]
\centering
\caption{Comparison between Model Settings}
\resizebox{\columnwidth}{!}
{\begin{tabular}{lcccc}
    \toprule
    &\multicolumn{4}{c}{Performance Metrics}\\\cmidrule{2-5}
    Algorithm&AUC&Precision&Recall&F-Score\\
    \midrule
    Random & 0.973 & 0.821 & \textbf{0.633} & 0.715 \\
    Averaging Embeddings & 0.971 & 0.816 & 0.628 & 0.71 \\
    LSTM + Attention & 0.976 & 0.877 & 0.62 & 0.727 \\
    \textbf{DAN} &  \textbf{0.978} & \textbf{0.888} & 0.628 & \textbf{0.735} \\
  \bottomrule
  \end{tabular}}

\label{tab:comparison}
\end{table}

Moreover, Table \ref{tab:uplift} shows the performance improvement to the Traveler Booking Intent (TBI) model when the Deep Average Network generated traveler embeddings are concatenated to the initial hand-crafted features.

\begin{table}[H]
\centering
\caption{Performance Uplift to TBI Model}
\resizebox{\columnwidth}{!}
{\begin{tabular}{lcccc}
    \toprule
    &\multicolumn{4}{c}{Performance Metrics}\\\cmidrule{2-5}
    Settings&AUC&Precision&Recall&F-Score\\
    \midrule
    Only Hand-Crafted Feat. & 0.975 & 0.817 & \textbf{0.651} & 0.724 \\
    \textbf{Hand-Crafted + DAN Feat.} &  \textbf{0.978} & \textbf{0.888} & 0.628 & \textbf{0.735} \\
  \bottomrule
  \end{tabular}}
\label{tab:uplift}
\end{table}

We noticed that the Deep Average Network traveler embeddings have competitive predictive power compared to the hand-crafted ones in the downstream TBI model. Based on random re-sampling the dataset and re-running the pipeline, we find that our results are reproducible.

\section{Conclusion}

We presented a method that combines deep and shallow neural networks to learn traveler and listing embeddings for a large online two-sided vacation rental marketplace platform. We deployed this system in the production environment. Our results show Deep Average Networks can outperform more complex neural networks in this context. There are various avenues to extend our study. First, we plan to test attention network without LSTM. Second, we plan to infuse other contextual information into our model. Third, we want to build a scoring layer that combines traveler and listing embeddings to personalize recommendations. Finally, we plan to evaluate numerous spatio-temporal features, representational learning approaches, and bidirectional recurrent neural networks in our framework.

\section{Acknowledgments}

This project is a collaborative effort between the recommendation, marketing data science and growth marketing teams. The authors would like to thank Ali Miraftab, Ravi Divvela, Chandri Krishnan and Wenjun Ke for their contribution to this paper.

\bibliographystyle{ACM-Reference-Format}
\bibliography{embeddings_recsys_2019}

\end{document}